%% file: main.tex
\documentclass[letterpaper,twocolumn,10pt]{article}
\usepackage{usenix2020_SOUPS}

\usepackage{tikz}
\usepackage{amsmath}
\usepackage{booktabs} % For formal tables
\usepackage{enumerate}
\usepackage{multirow}
\usepackage[utf8]{inputenc}
\usepackage{color, colortbl}
\usepackage{tabularx} 
\usepackage{dirtytalk}
\usepackage{balance}
\usepackage{graphicx}
\usepackage{pdfpages}
\usepackage{hyperref}
\usepackage{appendix}
\usepackage{xcolor}
\definecolor{Gray}{gray}{0.9}
\usepackage{xcolor}
\begin{document}
%-------------------------------------------------------------------------------

%don't want date printed

% make title bold and 14 pt font (Latex default is non-bold, 16 pt)
\title{\Large \bf Security, Availability, and Multiple Information Sources: \\Exploring Update Behavior of System Administrators}

% if you leave this blank it will default to a possibly ugly attempt 
% to make the contents of the \author command below into a string
\def\plainauthor{Tiefenau et al.}

%for single author (just remove % characters)
\author{
{\rm Christian Tiefenau}\\
University of Bonn
\and
{\rm Maximilian H\"aring}\\
University of Bonn
\and
{\rm Katharina Krombholz}\\
CISPA Helmholtz Center for Information Security
\and
{\rm Emanuel von Zezschwitz}\\
University of Bonn, Fraunhofer FKIE
}% end author

\maketitle
\thecopyright

\input{data.tex}

%-------------------------------------------------------------------------------
\begin{abstract}
%-------------------------------------------------------------------------------
Experts agree that keeping systems up to date is a powerful security measure. Previous work found that users sometimes explicitly refrain from performing timely updates, e.g., due to bad experiences which has a negative impact on end-user security. Another important user group has been investigated less extensively: system administrators, who are responsible for keeping complex and heterogeneous system landscapes available and secure.

In this paper, we sought to understand administrators' behavior, experiences, and attitudes regarding updates in a corporate environment. Based on the results of an interview study, we developed an online survey and quantified common practices and obstacles (e.g., downtime or lack of information about updates). The findings indicate that even experienced administrators struggle with update processes as the consequences of an update are sometimes hard to assess. Therefore, we argue that more usable monitoring and update processes are essential to guarantee IT security at scale.
\end{abstract}

\section{Introduction}
\label{sec:introduction}
``Keep your systems up to date'' is one of the most popular pieces of advice that security experts give to end-users~\cite{ion2015no,Reeder2017}. Supporting this, Khan et al. found that there is a correlation between not deployed updates and infected machines ~\cite{Khan2012}. Systems can easily be hardened against vulnerabilities like Heartbleed
\footnote{\url{http://heartbleed.com/}, accessed 02/25/2020.} by applying updates. Regardless of that, many systems in the wild remain vulnerable for two years or more~\cite{Symantec2016}. A prominent example of a situation where an update could have prevented severe data leakage is the Equifax breach \footnote{\url{https://www.theverge.com/2017/10/3/16410806/equifax-ceo-blame-breach-patch-congress-testimony}, accessed: 11/20/2019.}, which occurred in 2017. Similar incidents seem not unusual as is reported by an industry report ~\cite{industryReportServiceNow}.

Related work studied user perceptions and experiences with system updates and found that the results are often not in line with current recommendations of experts from a security perspective. In most cases, concerns about functional issues or unexpected UI changes hinder individuals from updating their systems~\cite{vaniea2014betrayed}.
In addition, users often do not understand the importance of non-visual changes~\cite{vaniea2014betrayed}, as they come with security updates. In contrast to users who are responsible only for managing their own personal devices, system administrators are in charge of large and complex IT infrastructures while also being users. We argue that their update behavior can have severe implications at a much larger scale. Marconato et al.~\cite{marconato2012security} observed the vulnerability life-cycle on different platforms and found that the time to patch and disclose vulnerabilities is decreasing. This finding can be applied to the Equifax breach and suggests that administrators are required to react in a timely manner.

Although general user concerns about system updates have been investigated in user studies, little light has been shed on the perspective of specific user groups (e.g., administrators or operators). 
Investigating administrators, Dietrich et al.~\cite{dietrich2018investigating} found that insecure configurations are often caused by institutional and individual factors, as well as time constraints. We assume that similar factors can have a negative impact on update processes. Administrators are often overworked \cite{dietrich2018investigating}, and updates are time-consuming. Secure systems, however, rely on updates and therefore, require regular attention by administrators.
As the body of literature is still in an early state regarding administrators' update behavior, we follow an inductive approach to explore the processes and obstacles that administrators face when updating in a corporate context. 

Our contributions are as follows:
\begin{itemize}
    \item We conducted \textbf{\qualitativeParticipantsAmount{} qualitative interviews to explore how administrators experience, perceive, and act during the update process}.
    \item We conducted an \textbf{online survey with \qntParticipantCount{} valid answer sets to test our observations on a larger scale.}
    \item We confirm that current \textbf{update processes and system factors tend to endanger IT security} and we discuss critical factors that need to be addressed to \textbf{support administrators}.
\end{itemize}

The results suggest that the update process differs among companies for various reasons. Administrators face a variety of obstacles in their update routines, e.g., downtime or hard-to-foresee situations, that often hinder them from performing updates in a timely manner. Overall, we argue that system administrators would benefit from more usable mechanisms, and that providing such mechanisms could effectively improve IT security at scale.

\section{Related Work}
\label{sec:relwork}
Two areas of research are specifically important for our work: 1) studies about update behavior and 2) studies investigating the security behavior of expert users. In the following, we present lessons learned from both research areas.

\subsection{Users' Update Behavior}
According to security experts, keeping systems and software up to date is an important security recommendation \cite{Reeder2017}. However, users may not follow this advice for reasons that are not related to security~\cite{redmiles2016think}, and only a minority of non-experts actually considers software updates an important security measure \cite{ion2015no, nicholson2018introducing}. It has been repeatedly shown that users often delay or even avoid updates \cite{forget2016or, moller2012update, Vitale:2017:HCS:3025453.3025509}. 

Investigation of the root causes of such critical user behavior has become a very active field of research. Previous work revealed diverse reasons for avoiding updates. Many users think that updates are not important because the link to security aspects often is not obvious \cite{fagan2016they,FURNELL20165, mathur2018quantifying, redmiles2016learned, Wash2014, vaniea2014betrayed,Vitale:2017:HCS:3025453.3025509}. Furthermore, users are often afraid of functional changes (e.g., UI modifications) \cite{bergman2017cognitive, vaniea2014betrayed, DBLP:conf/chi/VanieaR16, Vitale:2017:HCS:3025453.3025509} or fear making mistakes \cite{forget2016or}. Inconvenience is an important factor as updates can cause interruptions and take time \cite{mathur2018quantifying,Wash2014, Vitale:2017:HCS:3025453.3025509}. Finally, bad experiences with previous updates and negative online reviews hinder the installation of future patches \cite{fagan2016they,mathur2018quantifying, tian2015supporting, DBLP:conf/chi/VanieaR16}. This problem seems self-perpetuating, because the frequency of security updates is influenced by the emergence of novel attacks and thus, cannot be controlled by the vendor alone \cite{sarabi2017patch}. However, high update frequencies can lead to further negative reviews \cite{FLEISCHMANN201683, potharaju2017longitudinal}.

Several countermeasures for mitigating the problem of delayed updates have been proposed. As one straightforward solution, automatic updates \cite{Wash2014} and silent updates \cite{duebendorfer2009silent, sarabi2017patch} have been deployed. Although such mechanisms are very effective in keeping software up to date, they often cause confusion and irritation as they hamper the user's understanding of what is happening on their machines \cite{Edwards2008, Wash2014}. Furthermore, some users might have good reasons to refrain from performing certain updates \cite{Edwards2008}. Therefore, user-centered solutions, such as providing more information \cite{mathur2016they, parthesarathy2002software, tian2015supporting, tian2014study} and designing better notifications \cite{FAGAN2015504, Fagan2015, Frik2018}, have been repeatedly suggested as complementary concepts to further increase compliance rates. 

\subsection{IT Professionals and IT Security}
Recently, researchers have started focusing on security-related usability problems of specific user groups~\cite{acar2016you}. In contrast to security advocates \cite{haney2018s} or security analysts \cite{Goodall2004}, most of these people are not security professionals. They are often knowledgeable in a specific domain, related to IT. Several recent studies addressed the problems of software developers \cite{Acar2016, acar2017developers, assal2018security, LeoGorski2018}. For example, Acar et al. \cite{Acar2016, acar2017developers} investigated available sources of information and how these sources influence code security. Gorski et al. \cite{LeoGorski2018} showed that software developers benefit from API-integrated security recommendations and that such usability-optimized concepts can significantly improve security \cite{Tiefenau:2019}. 

Several human-centered studies with system administrators were published between 2001 and 2007. In 2001, Hrebec and Stiber \cite{Hrebec2001} studied the mental models of system administrators and found that these experts often struggle to understand the complex systems that they need to manage. In addition, the study participants reported a lack of formal education and the desire to solve problems by themselves. Barrett et al. \cite{Barrett2004} found that system administrators often lack situational awareness.
Haber and Kandogan \cite{Haber2007, kandogan2005security} and Botta et al. \cite{botta2007towards} observed the tools and work practices of security administrations and IT professionals. Their results show that security administrators perform a lot of different tasks and need various skills like pattern recognition or inferential analysis to perform these tasks. They proposed, that new classes of tools need to be developed to counter the ever increasing complexity of the systems and attack-vectors.

In contrast to this early work, a few recently published papers investigated more specific problems of system administrators. Fahl et al. \cite{fahl} studied non-validating X.509 certificates and revealed that about 30\% of the responsible webmasters misconfigured their web servers accidentally. Ukrop et al. \cite{2019-acsac-ukrop} analyzed the corresponding warnings and found that rewording can help administrators to make better informed decisions. Krombholz et al. \cite{httpskrombholz,krombholz2019if} showed that the deployment process for HTTPS is far too complex and that administrators struggle with finding secure and compatible configurations due to the lack of conceptual mental models. Dietrich et al. \cite{dietrich2018investigating} investigated the administrators' general perception of misconfigurations and identified missing or delayed updates as one of the root causes of these problems. 

There exists work that discussed update processes in companies \cite{BLYTHE201887, blythe2015unpacking, min2007deciding, Vitale:2017:HCS:3025453.3025509}. For example, Vitale et al. \cite{Vitale:2017:HCS:3025453.3025509} performed three interviews with technical staff concerned with updates and found that these professionals prioritized security aspects and licensing issues over potential usability consequences. This finding confirmed previous findings \cite{min2007deciding} that in a corporate context, business needs rather than user requirements drive update decisions. In contrast, Blythe et al. \cite{blythe2015unpacking} reported that employees often rely on ``security experts'' in the company to manage updates and often lack a feeling of responsibility. Finally, the update challenges of system administrators have been indirectly considered by various researchers who proposed automatic tools to improve the manageability of the update process (e.g., \cite{bachwani2012mojave, gallagher2014verifying, latimer2017trace, oberheide2009if}). However, none of these concepts have been evaluated in a user study.  

Parallel to our work, Li et al.\cite{Li:2019} published a closely related paper in which they studied US-based system administrators in a qualitative fashion. They as well researched the update process in companies and found several \textit{pain points} within the process. In contrast, our interview sample was drawn from German companies, thus representing a different culture. Overall, our study confirms most of their findings. We will separately discuss our findings in comparison to Li et al.'s in section~\ref{sec:discussion:li} in more detail.

\section{Interview Study}\label{sec:qual}
Although recommendations for patch management have been published\footnote{\url{https://www.infosec.gov.hk/english/technical/files/patch.pdf}, accessed 02/25/2020.}, we are aware of only one other study that systematically investigated the update behavior of system administrators\cite{Li:2019}. Therefore, we started with an interview study to identify important factors of the problem space. 

This interview study aimed to provide answers to the following research questions with an emphasis on administrators' perceptions, challenges, and tools they use in their update routines:

\begin{enumerate}
    \item \textbf{How can the update processes be described, and what common patterns are there?} \\
    Administrators are usually paid professionals who are responsible for updating large and complex IT infrastructures. This raises the question, whether, and if so, where, system administrators' updates processes differ from end-users' processes\cite{DBLP:conf/chi/VanieaR16}.

    \item \textbf{What issues and obstacles do professional administrators face in their update routines?} \\
    We specifically aim at understanding the problems of administrators and their perception of update processes. Identifying obstacles in relation to processes, tools, and environments is indispensable to define important directions for future work.

    \item \textbf{How are administrators informed about updates, and which sources of information do they use?} \\
    Related work has indicated that the source of information can have a significant impact on software security \cite{Acar2016, fischer2017}. Thus, we aim at understanding how administrators gather information and what sources they use.

    \item \textbf{What kind of tools do administrators use to manage system updates, and is there room for improvements?} \\
    As usable security researchers, we are specifically interested in the tools involved in the update process. We hypothesize that although some tools are used on purpose and other tools are unavoidable, such tools can either complicate or ease the process. 
\end{enumerate}

\subsection{Study Design and Procedure}
We conducted \qualitativeParticipantsAmount{} semi-structured interviews in June 2018 to explore the participants' opinions, thoughts, and experiences.
Based on three pilot-study interviews, we refined the interview guidelines to balance between informing the research questions and supporting a flexible exploration of the problem space (i.e., leaving enough room to add further comments). 
The interview was structured into (1) general questions about the daily work routine of the participant, (2) general experiences with updates, (3) a more detailed assessment of specific aspects, and (4) additional comments. The guidelines are in Appendix~\ref{app:interview_guideline}.

All but one interview were conducted by the same researcher. Both researchers are experts in computer science and spoke the same native language as the interviewees. After an introduction to the purpose of the study, the participants were asked to sign a consent form. All participants gave their consent to being audio-recorded. We conducted one interview in person and six via telephone. All interviews were held in German. During the interviews, the interviewee and the researcher were allowed to take notes. The interviews lasted between 34 and 67 minutes and ended with a short questionnaire that collected demographic information.

\subsection{Recruitment and Participants}
\label{recruitmentandparticipants}
We did not restrict our invitations to administrators working with a specific operating system, infrastructure or type of update. The only criterion for inclusion was that participants had to be in charge of, or in contact with, any kind of updates. Personal contacts were used as entry points to larger organizations and asked to forward the announcement to their employers' IT department. Additionally, we directly approached representatives of medium-sized and large companies at CeBIT 2018, a large international computer expo\footnote{\url{https://www.cebit.de/}, accessed 02/25/2020.}. 

In total, we recruited \qualitativeParticipantsAmount~participants at companies that had an office based in Germany. All participants reported they were in charge of system administration, although they had various job descriptions and managed different types of systems. Table~\ref{tab:interview_participants} in the appendix presents more details about the sample. All the participants were male. For ease of readability in the following sections, we assigned the participants random names.

\subsection{Analysis}
\label{analysis}
The interviews were transcribed, and coded by two researchers. We coded open answers inductively following the approach of Wertz, Charmaz et al. \cite{Wertz}. The two researchers categorized the data according to the research questions presented in Section~\ref{sec:qual}. The first three interviews were coded in a batch to establish the first codebook. Each of the following four interviews was coded separately. Then, the conflicts were discussed, and new codes were added to the codebook.
We calculated the combined Krippendorff's alpha \cite{Krippendorff2004} before (\alphaBefore) and after (\alphaAfter) the discussion phase for each interview. Our goal was to use the qualitative analysis solely as a first step and foundation for the following quantitative study. Therefore, we refrained from continuing with interviews until theoretical saturation \cite{Francis2010} was reached.

\subsection{Qualitative Results}
\label{sec:qual:preliminaryresults}
\input{tables/qual_phases.tex}
In the following, we present the results from the interview study with respect to the research questions. 

\subsubsection{Update Processes}%Q1
\label{sec:qual:process}
In Table~\ref{table:qual:phasesandsteps}, we present the sum of all extracted process phases, including all reported steps that were performed in these phases.
Overall, the update process varied in time and structure among participants and tended to be variable even for individual administrators, depending on the software that needed an update.
Cyril reported he worked in a client environment with Windows systems. He was concerned mainly with regular update cycles. Therefore, he was able to prepare for update events (e.g., briefing the team, allocating resources, allocating maintenance windows, and gathering information). Four out of seven participants reported they relied on fixed update cycles for client systems, although Zelko reported that this was not always possible in practice. In contrast, Lorenz, who worked at a smaller company, reported that employees at his company were responsible for their systems. When we discussed more specific software, the answers became more diverse. Milan usually builds packages to automate the distribution, but Markus tends to perform manual installations.

Although participants' responsibilities differed, we were able to identify common patterns in the update process. Most of these phases can be mapped to those of client users \cite{DBLP:conf/chi/VanieaR16}. However, we identified three major differences:

\textit{Some administrators perform extensive testing} before installing the update on a live system. For example, Julian utilized up to three stages. Zelko, who stated, that \say{[E]ven if there is a risk that the update breaks something, we install them timely}, utilized two test stages. First, he tested the update with virtual machines that simulate the client landscape, and then he rolled out the updates for a small group of colleagues. 

\textit{Updates are rolled out step by step}. The participants reported that often not all systems are updated in one batch. This allows the administrators to minimize the number of misconfigurations once an update fails, but constraints on resources are also a reason for this. For example, Julian reported that the network would be used to capacity if all systems were patched at the same time.

\textit{The preparation step is structured and involves planning and research} of resources and the allocation of time slots. Five participants explicitly reported they conduct online research before they install an update. In addition, Alexander told that important update decisions are often made in group discussions.

\subsubsection{Obstacles}
\label{sec:obstacles}
We identified various obstacles that hamper the administrators' task of performing updates. In Table~\ref{table:qual:phasesandsteps}, we connect and
report obstacles to the phases of the update process.
In the following, we discuss common obstacles in more detail:

\textit{Downtimes.}  
The participants stated that downtimes are a serious obstacle in the update process which often cause delayed deployments. As soon as a reboot is necessary, and there is no redundant system, downtime is induced. Alexander gave anecdotal evidence of a mitigation strategy: Upgrading from Solaris 10 (which required significant downtime) to Solaris 11 (which supports near to hot-swap updates and an easy rollback) increased update frequencies from three times a year to once a week. 
	
\textit{Dependencies.}
The participants reported patches that break dependencies usually delay the process. Although this may not be surprising, it highlights the problem of dealing with dependent systems that cannot be patched in time. Further dependencies are introduced as part of the infrastructure landscape. For example, some systems depend on other systems to be available at boot time (Markus). Assessing these dependencies and then following the right order makes the process highly complex.
Another type of dependency is towards the vendor of the software or hardware. An example of this can be as trivial as no available patches, even if a vulnerability is public, as Lorenz reported for the Meltdown case. 

\textit{High frequency and large files.} 
Every update takes resources: for example, time, workforce, CPU, and data storage. Zelko reported that big update files, which are often a consequence of combining functional updates with security patches, can cause problems. To handle resource constraints, updates are rolled out in multiple but smaller batches (Julian). 

\textit{Competing priorities.} 
Similar to standard users, administrators' decisions to perform updates are influenced by various factors. Participants reported stability considerations, the risk of an exploit, and performance issues as influential aspects. The fact that some systems do not separate security and feature updates may intensify this situation. Finally, required resources are sometimes allocated to other processes that have higher priority. Alexander reported that \say{the decision [to update] is always based on the sum of available information}. As mentioned in Section~\ref{sec:qual:process}, group discussions are an important part of the process. However, the need for communication can also delay updates (Milan).

\textit{Human Factors.} 
In addition to technological and structural constraints, the administrator faces other obstacles. 
Missing expertise or a lack of knowledge can lead to situations where administrators rely on third parties. In this regard, Lorenz acknowledged that he does not always know how to act correctly. Or as Markus put it, he has to trust the vendor that the classification of the patch is correct.
System administrators have to trust the information they get from the software developer, vendor, or other source. Another factor we identified is social pressure, as Lorenz reported, \say{And you look like an idiot, when you kill a git server. [...] That chases me.} 
Another aspect that makes updating harder for administrators was software which is managed by end-users. Such software is often installed without the knowledge of administrators and makes the update process more complicated because it is not integrated in standard processes.

\subsubsection{Sources of Information}
The participants reported they use various methods to inform themselves about security updates and vulnerabilities. Five out of seven participants reported they use third-party sources that were independent of the software publisher, such as popular news portals or blogs. This information is usually supplemented by publisher-related newsletters and specific mailing lists, such as the Ubuntu-security mailing list (Lorenz).
Cyril mentioned specialized third-party services that push information about available patches.
Others got more specific and reported that they use tools like SCCM\footnote{\url{https://www.microsoft.com/en-us/cloud-platform/system-center-configuration-manager-features}, accessed 02/25/2020.} or Nessus\footnote{\url{https://www.tenable.com/products/nessus/nessus-professional}, accessed 02/25/2020.} which serve as sources of information. 

\subsubsection{Tools}
\label{qal:res:tools}
The participants reported OS-integrated tools and special purpose tools that are used to update servers and clients and that serve as sources of information. The purpose of such tools ranged from monitoring systems (Julian) to complete automation of the update process, such as SCCM or WSUS\footnote{\url{https://docs.microsoft.com/en-us/windows-server/administration/windows-server-update-services/get-started/windows-server-update-services-wsus}, accessed 02/25/2020.} (Markus). Participants also named external services (e.g., Shavlik\footnote{\url{https://www.ivanti.com/company/history/shavlik}, accessed 02/25/2020.}) that test and pre-filter patches for companies. Although automation of update processes was an important goal for participants, it had not yet been fully implemented. Software that is not covered by such tools, meaning not integrated by default, has to be updated manually or integrated. This seems to be the case when the vendors or the operating systems differ (e.g., using Microsoft WSUS to update Adobe Flash Player). Although the integration is possible, it is connected to additional effort and is not always done (Markus), e.g., if it affects only a small group of clients (Milan). 
Concerning future developments, Lorenz was less optimistic and brought up that the time investment in tools that would ease the workflow was not a high priority.

\subsection{Key Observations}
\input{tables/quant_observations.tex}
\input{tables/quant_eval.tex}

We performed an interview study of administrators' update behavior. Based on the research questions, we were able to describe update processes, common obstacles, information retrieval, and the use of software tools.
We extract a series of key observations to guide the construction of the quantitative study, following the interviews. Table~\ref{table:qual:phasesandsteps} provides an overview of the process phases, tools, and obstacles that administrators face in their daily lives according to the participants.
Table~\ref{tab:quant_observations} presents nine key observations, which were formulated based on the qualitative findings and then categorized in three groups: ``Update Process and Information,'' ``Update Obstacles,'' and ``Human Factors.'' 
In the next section, we report on a quantitative online survey which was performed to shed further light on the update behavior of system administrators.

\section{Quantitative Online Survey}
\label{sec:quant}
Following the interviews, we performed a quantitative online survey. We created statements based on our observations in the interview study and developed an online questionnaire to quantify and enrich them. 

\subsection{Procedure and Structure}
The recruitment process for the preliminary interview study indicated that system administrators are inherently short on time, and thus, minimizing the time to fill out the survey was indispensable to obtain a sufficient number of responses. Therefore, most of the questions were based on simple answer types, such as check boxes or rating scales. To further motivate participation, we offered an opt-in for a raffle of 3D prints. Every tenth participant had the chance to win a 3D-printed model of their choice. E-mail addresses were collected only for this raffle, stored separately, and deleted afterward. Twenty-three entered their contact email address of whom no one was interested in a print. After participants had given their consent to take part in the study, the survey started. Completion took about 10 minutes.

To support many different circumstances, we framed questions in a way that answers could be related to the current position or if not applicable, to the last position as system administrator. We started by collecting demographic data (e.g., age), personal information (e.g., years of experience), information about the work environment (e.g, their role, company size), and information about update processes (e.g., existence of formal processes). In the second phase, participants rated 1) the frequency of specific events using 5-point scales ranging from ``1 - Never'' to ``5 - Always'' and 2) indicated their agreement with different statements using 7-point scales (``1 - Strongly disagree'' to ``7 - Strongly agree''). The questions were presented in random order for each participant. The questions were chosen based on our observations and thus, examined the impact of obstacles (e.g., ``Downtimes caused by the update process hinders the installation of an update''), human factors (e.g., ``I feel that I am sufficiently trained as an administrator''), and information sources (e.g., selection of sources used). The questionnaire ended with an open-ended question about the biggest obstacles in the update process that we coded afterwards. The new categories are marked with an asterisk in Table \ref{table:qual:phasesandsteps}.

To ensure the internal consistency of the collected data, we added an attention check based on the negation of one of these questions. Five participants, who answered both questions with a different polarity, were excluded from the evaluation.

\subsection{Recruitment and Participants}
To attract professional system administrators, we decided against using crowdsourcing platforms like Amazon Mechanical Turk. Instead, we reached out to community sites like Reddit and specialized forums. Additionally, we used Twitter and followed a similar approach as we did in the interview study. Posting in forums resulted in 66 answers, advertising on Twitter resulted in 67 responses, and using personal contacts in companies to spread the questionnaire contributed eight answers. 

The English survey was active for 14 days in September 2018. During this time, the questionnaire was started 141 times and completed by 72 ($51.1\%$) participants. As reported, five data sets were excluded from the analysis due to failed attention checks, resulting in 67 valid data sets. The participants' age ranged between 22 and 55 years. Fifty-eight of them were male, one female, three reported ``Other'' and five preferred did not specify their gender.
More than $61\%$ ($41$) work in European countries. The biggest group of the participants pool work in Germany ($22$), but we also received answers from other continents, like North America($19$), Australia ($2$) or South America ($1$). Table ~\ref{tab:quant_demographic} in the appendix provides an overview of the participants' demographics.
The job-related education of our participants can be classified as ``unspecified training,'' ``vendor training,'' ``self taught,'' and ``experience at the job.'' Most of the participants worked in a team (39), 16 were a team leader, and 10 worked alone. In the following, we report on the data gathered by the questionnaire.

\subsection{Results}
In the following, the results of the online survey are presented structured by the main categories presented in Table~\ref{tab:quant_observations}.
The observations from the interviews suggest that company size may have an influence on different factors. To assess this point, we divided the data sets in two groups: 34 companies with 250 employees or fewer were tagged as small and medium-sized enterprises (SMEs), and 33 companies with more than 250 employees were defined as large enterprises \cite{eu2003smedefinition}. 
This was found to be a suitable comparison because post-hoc we had comparable group sizes. A controlled analysis of additional factors was not feasible at this stage, and future work should consider other aspects (e.g., experience, type of systems, and team size). 
Tables \ref{table:quant:statementresponsesfivepoint} and \ref{table:quant:statementresponsessevenpoint} show the answers of the participants to the statement they were presented. 

\subsubsection{Update Process and Information}
\begin{figure}
\includegraphics[width=\columnwidth]{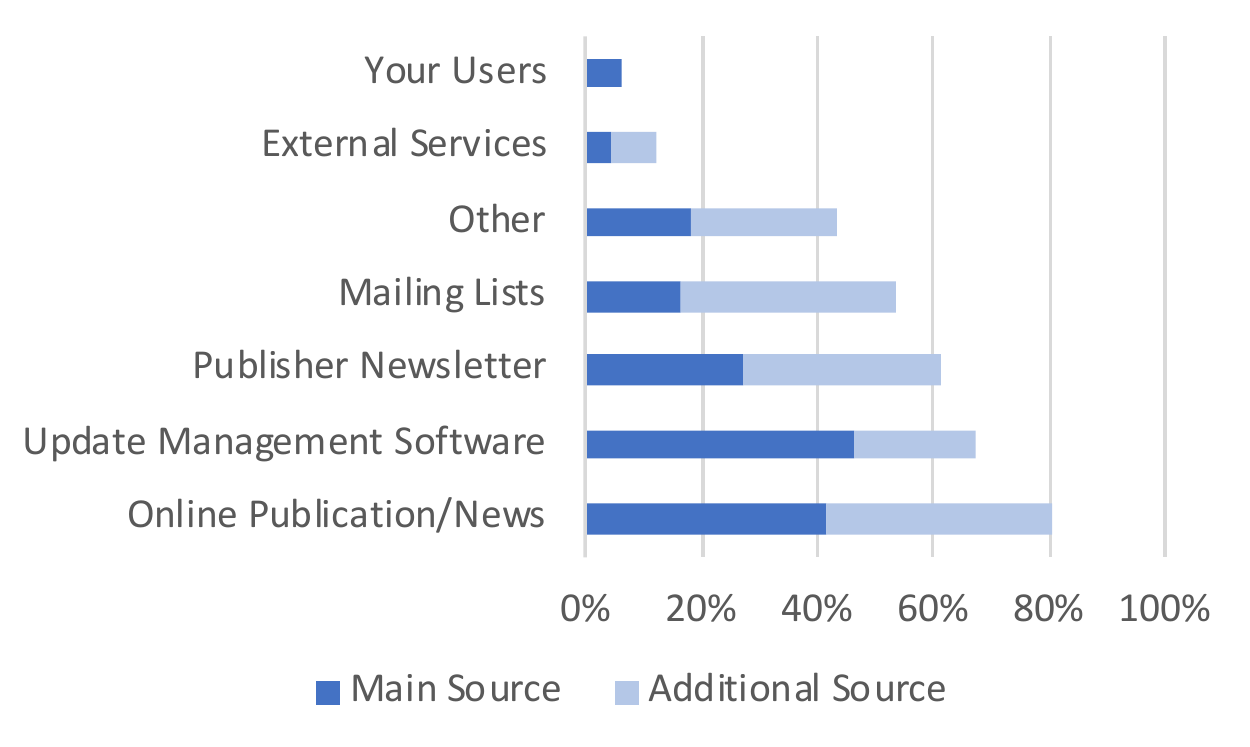}
\caption{Distribution of information sources used by the administrators (n=67).}
\label{fig:quant:sources}
\end{figure}

\paragraph{U1}
Figure \ref{fig:quant:sources} presents the \emph{sources of information} administrators use to learn about (new) updates. Most of the participants reported a median of three different sources. Third-party online publications are the most frequently used sources of information. They served as a source for 54 (81\%) participants, and 28 of all 67 participants (42\%) even declared them the main source of information. When focusing on the main source of information, we found that update management tools are essential for most administrators (46\%). Fisher's exact test indicated no statistically significant differences between differently sized companies ($p = 0.2242$). Using an optional comment field, some administrators added other sources of information, such as vendors, the online community (e.g., Twitter), work experience, and active monitoring of systems. Due to the structure of the questionnaire, we cannot make statements about how the participants ranked the quality of those sources. We do not know whether they use one source to get informed about the occurrence of an update and then use another to capture details. 

\paragraph{U2}
To investigate the \emph{existence of formal update processes}, we asked the participants if 1) ``there is a written document,'' 2) ``no document but an informal guideline,'' or 3) ``no defined process'' in their company. Twenty-eight (42\%) participants indicated the existence of formal processes, 26 ($39\%$) administrators had at least informal guidelines for performing updates, and 13 ($19\%$) participants indicated that there are no predefined processes. A comparison of the use of formal, written update processes in differently sized companies revealed a statistically significant difference between large companies ($57.6\%$) and smaller ones ($26.5\%$), ($p = 0.0136$, $ratio = 3.769$, Fisher's exact test). This indicates that small companies make less use of formal update processes. The lack of such a process is not uncommon in our sample, as 10 out of 34 of the small companies did not report any kind of defined process.

\subsubsection{Update Obstacles}\label{sec:obstaclesProcess}
\begin{figure}
\includegraphics[width=\columnwidth]{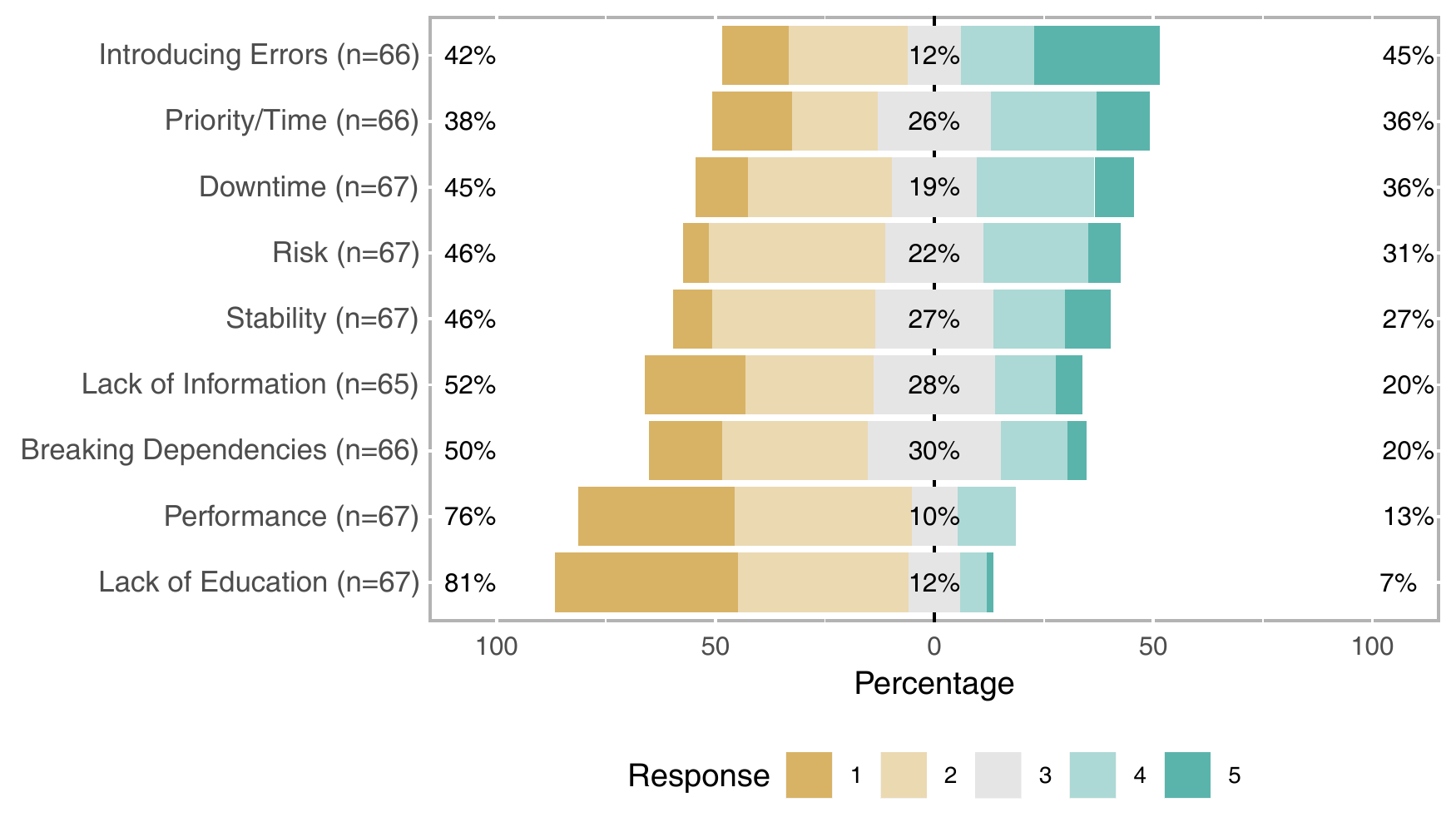}
\caption{Frequency of considerations that hinder the installation of an update. The scale ranged from ``1 -- Never'' to ``5 -- Always.'' Not included are ``not sure'' or missing answers.}
\label{fig:quant:obstacles}
\end{figure}
Figure \ref{fig:quant:obstacles} shows the share of administrators who have faced specific obstacles during daily update routines. Quantifying the observations, we found that general risk assessments are known to most of the participants ($94\%$) while deciding to deploy specific updates. Only four ($6\%$) participants answered that they never considered assessing risks as an obstacle, while $63$ agreed they did so at least sometimes.

\paragraph{O1 to O4}
When asked about more specific obstacles, or risks, \emph{stability considerations} represented the biggest issues that had been considered by 61 ($91\%$) participants in the past. Similarly, 59 (88\%) participants considered \emph{downtime} as a specific obstacle. \emph{Lack of information} ($50, 77\%$), performance issues ($43, 64\%$) and educational aspects ($39, 58\%$) were the least prevalent obstacles in the sample. However, even those factors were considered by a majority of the participants. Finally, we performed Mann-Whitney U tests to investigate the impact of company size on the prevalence of obstacles: We could not find statistically significant differences \footnote{stability considerations: $p=0.814$, downtime: $p=0.324$, lack of information: $p=0.655$, performance issues: $p=0.067$, educational aspects: $p=0.752$, introducing errors: $p=0.611$, risk considerations: $p=0.415$, breaking dependencies: $p=0.387$, priority: $p=0.559$}. 

Fifty-five percent seemed to agree that \emph{problems after the installation of an update} are only a minor concern. However, eight participants strongly disagreed with the statement. Five were undecided, and 25 ($37\%$) disagreed in some way. To cover potential reasons for the answers, we assigned participants to two groups: those who do some kind of testing before installing updates on the live system ($n=45, 67\%$) and those who do not ($n=22, 33\%$). There was no statistically significant difference ($p = 0.2553$, Mann-Whitney U test) meaning that having a testing phase seems not to prevent all problems after the installation. Due to the sample size, we could not investigate if the company size is a significant factor in this regard.

\paragraph{O5}
Another aspect in the interview study was the \emph{user rights}. The agreement to the observation “Users often install software without the administrators' knowledge” was diverse. Although there was a tendency to disagree, as can be seen by the low median ($3$), there were also seven strong agreements. We found no statistical significance that would have supported our assumption that IT companies may have a different distribution on this than non-IT companies. 

\subsubsection{Human Factors}
\paragraph{P1}
Seventeen (25\%) administrators reported that they always \emph{feel sufficiently trained} for dealing with updates. However, 50 ($74.6\%$) participants already faced situations for which they did not feel sufficiently trained. An evaluation of the impact of the administrator's company size indicates that administrators at large companies ($Median=4$) more often feel sufficiently trained than their colleagues at smaller companies ($Median=4$), Mann-Whitney U test: $U=358.0, p < 0.01$, two-sided.

\paragraph{P2}
\input{tables/quant_timespan.tex}
Finally, all administrators (except one) somewhat agreed that \emph{timely updates are important}. The self-reported time span between the release of an update and its installation can be seen in Table~\ref{tab:quant:timespan}. While some participants reported deploying updates within a day, there were nine cases where updates needed more than a month.
Optional comments given by the participants supported the findings that downtime, complexity, and dependencies are common reasons for such delays.

\subsubsection{(Missing) Distinction between Security- and Feature-Updates}
The interviews revealed that security- and feature-updates are often hard to distinguish. While we did not ask for the share of security-related patches in our interviews, the survey participants reported that 56\% (ranging from 5-100\%) of the overall updates involved security-related ones.

\section{Discussion and Implications}
Our work identified multidimensional problems that should be addressed by multiple stakeholders (e.g., software vendors or the companies themselves).
In this section, we reflect on our results, provide actionable recommendations for these stakeholders and suggest directions for future research. 
We acknowledge that many aspects reported in this paper may seem like "common sense". With this work, we add to the scientific evidence in this very broad area with several factors that influence the update process and directions for further research and discussion.

\subsection{Security Implications}
Our results are in line with Li et al. and show that even professionals cannot always deploy updates in a timely fashion. This can be a security issue since outdated systems are often vulnerable to exploits. The administrators we asked were aware of this problem and agreed that deploying updates in a timely manner is important. However, we found that external factors such as compliance with company-specific rules, inflexible processes and communication overhead (e.g., leadership approval) still delay updating in practice.
Future work needs to take a more holistic view and investigate technical and social factors in the update process. We need to understand which people are involved in these processes and how their communication can be supported. In addition, we need to develop approaches to better communicate the urgency of specific patches as today, the rating is often not clear\cite{industryReportServiceNow}.

\subsection{Update Process} 
The results showed that the update processes of system administrators are diverse and complex. Although the update processes of administrators can be matched to the end-user phases \cite{DBLP:conf/chi/VanieaR16}, the identified phases differ in the details. In particular, gathering information and discussing update decisions were identified as important but time-consuming steps. As many administrators reported they make decisions in group meetings, we raise the question of how individual administrators can be supported in their decision-making process. The preparation process takes time and involves extensive testing. Although the testing processes were handled differently, they usually involved multiple iterative stages. This indicates that administrators have to go through the whole update process multiple times. Two findings were primarily interesting: 1) Many companies lack formal processes, and 2) the update process is highly complex and lacks automation.
The insights into this process provide important directions for future research and immediate action items for software vendors, such as the following:
\begin{itemize}
    \item Formal processes seem to be more frequently used in large companies. Whether formal processes help to reduce the burden of decision-making and ease the overall process should be researched; that is, in what way they influence the update process (e.g., can well-defined responsibilities speed up the decision and do they lead to more and faster updates?) and where possible trade-offs can be expected (e.g., decreased complexity versus more time needed). 
    \item The high number of iterative steps must be supported, e.g., with automation approaches. Thus, it is important to understand which phases of the process are critical and which parts can be effectively supported by tools.
    \item A possible approach for improving the process could be to connect more effectively virtual teams of administrators who share similar responsibilities and manage similar systems. Supporting such concepts with feasible tools can quickly lead to shared knowledge of best practices and experiences resulting in a better overview of the effects updates have on their systems. We hypothesize that especially smaller companies would profit from that.
\end{itemize}

\subsection{Obstacles}
The findings indicate that administrators face severe obstacles that often hinder them from performing timely updates. In line with Dietrich et al.'s work~\cite{dietrich2018investigating}, the findings show that the problems administrators face are diverse and interconnected. 
Corresponding to Hrebec and Stiber's findings~\cite{Hrebec2001}, individual-related factors, such as negative and positive experiences with updating, as well as education, come into play.
The findings provide a baseline for future research questions and immediate action items for software vendors, such as the following:
\begin{itemize}
    \item Due to the highly diverse landscape of large-scale systems, future research should further explore contextual factors and different populations of administrators. Differentiation of the various types of administrators could help to better categorize participants and understand their diverse problems and challenges. Related to this point, the check of the external validity of the research would benefit from better differentiation of types of administrators. However, a practicable taxonomy for this is still missing.
    \item Software development should focus on reducing downtime and providing rollback mechanisms that encourage administrators to take the risk of potential negative effects on availability.
    \item Researchers and software vendors should investigate on how to provide reliable information and accurate documentation of the effects of an update and occurring problems right in the moment and at the place the update is going to be installed.
\end{itemize}
Therefore, we hypothesize that supporting administrators' situational overview will have positive effects on timely updates. Finally, minimizing consequences by providing reversible updates, or just updates that have very small effects, could furthermore help administrators to update.
As an example, dynamic software updates (DSU) \cite{HicksDSU} seems like a promising technique to contribute to this area and could be evaluated from this perspective.

\subsection{Coping Strategies}
As a consequence of facing obstacles, system administrators have developed a diverse set of coping strategies.
Although the degree of usage varied among participants, an important countermeasure against the growing complexity is the use of tools that monitor update processes and support to (partly) automate installations. Because administrators expressed the desire for more automation, the findings emphasize the importance of the area of research that deals with the development of such concepts \cite{bachwani2012mojave, gallagher2014verifying, latimer2017trace, oberheide2009if}.

To cope with the problem of limited resources combined with growing package sizes, the participants started to divide update processes into multiple batches. This can have the advantage of allowing more feedback loops and of reducing the load on the network. However, at the same time, this process increases the number of required iterations for single patches. Although we argue that the footprint (e.g., resources needed to roll out), especially of security updates, should be minimal, this may not always be possible.

Based on the findings, we provide the following recommendations to support existing coping strategies and for the development of novel solutions:
\begin{itemize}
    \item Hot swap functionality and small-sized patches which enable administrators to estimate the impact of the installation on their systems, have the potential to further ease the update processes.
    \item Update management tools should better support the integration of third-party software.
    \item Administrators' coping strategies are still not sufficiently understood. Thus, researchers should focus on systematically investigating different coping strategies for various obstacles, identify desirable behavior and analyze in which way the human aspect contributes to this.
\end{itemize}

\subsection{Comparison to Results by Li et al.~\cite{Li:2019}}
\label{sec:discussion:li}
As mentioned in Section~\ref{sec:relwork}, a thematically similar publication emerged independently while we were working on this research. Li et al. published a study on system update processes among US American system administrators, identifying an update process that was very similar to ours~\cite{Li:2019}. 

\subsubsection*{The Update Process}
While Li et al.'s process emerged entirely from their interview response data, our update process was informed by theoretical work by Vaniea et al.~\cite{DBLP:conf/chi/VanieaR16}. This could explain minor differences such as the separate testing phase we introduced to highlight the difference to the end-user process. Despite this difference, overall, we consider the identified processes to be very similar.
In alignment with their findings, we can confirm that in the information-phase, administrators use multiple sources to derive information about updates. We didn't find any statistical difference in the number of sources used between administrators working in different companies (big vs small) in our sample. Li et al. reports on the frequency of the used sources and that three quarter of their participants used security advisories or direct vendor notifications. In our data, $81\%$ informed themselves using online publications and $63\%$ relied on publisher newsletters. We can add that despite having multiple sources (median=3), our population uses update management tools as their main source followed by online resources.

Both works identified the deciding-phase. We can match most of our identified obstacles to the reported factors of Li et al. With a slightly different perspective, we can add an additional reported obstacle that focuses more on the administrator executing the process than the update: missing expertise. 

We can support Li et al.'s finding that testing is an important phase in the process and we encountered the same approaches: ``Staggered deployments'' and ``Dedicated testing environments''. As 83 of 102 (81\%) of their survey participants included some form of testing, a slightly smaller, but still the major, part of our participants 45/67 (67\%) reported the same.

As for the remaining two phases, our works differed in focus. While Li et al. extensively discussed the method of deployment (automatic vs. manual) and the decision of when to deploy in the deployment phase, our work concentrates on the obstacles the administrators face in this phase. For the post-installation phase, their work presents the ways in which administrators deal with update issues, while we report on the frequency of the occurrence of such issues (O3) in Section~\ref{sec:obstaclesProcess}. 

\subsubsection*{Obstacles in the Update Process}
Li et al. identified challenges faced by administrators within this update process that can be categorized as: (1) obtaining relevant information about relevant updates and deciding, (2) preparing, testing and deploying updates in a timely fashion, (3) recovering from update-induced errors, and (4) organizational and management influence~\cite{Li:2019}. Our identified obstacles (cf. Section~\ref{sec:obstacles}) are in line with these obstacles.
Li et al.'s work reports that identifying the relevant information in an update can be a challenging task. We can confirm this (O4) and show that this was mentioned by 77\% of our participants.

Automation can help to deploy updates sooner and more frequently. Li et al. have found several obstacles such as dependency and compatibility considerations or host heterogeneity as factors that have an influence on update deployment. In addition to those, we have found additional ones such as missing tools or performance considerations in our data set. Table \ref{table:qual:phasesandsteps} provides a summary of our findings that assigns the problems to the phases in which they occur.

In general, while their work reveals the existence of those problems, we can complement these problems with the frequency of the problems that our survey participants stated. Li et al. report that the recovery of updated-induced errors is a problem that we can enrich with the fact that this seems to be of mixed importance (O3). This could indicate that this is a context-dependent factor, and a more detailed research must be undertaken in this regard.

Also, Li et al.'s work reports on the existence of organizational oversight that hinders or delays updates in some cases. We can also find this problem and show that this, among stability and risk considerations, is of more importance than factors such as performance considerations.

\subsubsection*{Demographics}
While both Li et al.'s and our study are very similar in methodology, they differ in a key point: the recruited sample. Li et al. sampled only US-based administrators, while we recruited our interview-study population from Germany and our survey participants were mostly (41 of 67) European-based. Despite work culture in the US and Europe (e.g. in Germany ~\cite{Hedderich2010, Pethe, faist2000volume}) being distinctively different (stemming from cultural differences in education, law, and professional socialization, among others), both studies report similar findings. We are thus in the fortunate situation to not only have our methodology and findings independently validated within a close distance in time, but also to confirm that the phenomena we identified are relevant across both US and European system administrators.

On interpreting the independently compiled findings, we have an indication that the system administration process is not as susceptible to cultural differences (at least in Western societies) as other fields of work. This might be connected to the rather globalized nature of IT infrastructure. Both participant pools used similar software, e.g., SCCM or WSUS (cf. Section~\ref{qal:res:tools}). It is reasonable to assume that the technical challenges are similar.
Comparing both papers, we could not find any differences that originate in individual or organizational factors. If this can be confirmed in further studies within different countries such as China (the largest producer of IT hardware and systems\footnote{\url{https://www.mckinsey.com/~/media/mckinsey/featured\%20insights/china/china\%20and\%20the\%20world\%20inside\%20the\%20dynamics\%20of\%20a\%20changing\%20relationship/mgi-china-and-the-world-full-report-june-2019-vf.ashx}}), Estonia (the often considered ``most advanced'' country within the EU in terms of digital transformation\footnote{\url{https://www.wired.co.uk/article/estonia-e-resident}, accessed 11/21/2019.}), or Qatar (the largest economy in the Middle East according to GDP per capita~\footnote{\url{https://www.cia.gov/library/publications/the-world-factbook/rankorder/2004rank.html, accessed 11/21/2019.}}), this would significantly widen the recruitment possibilities for future studies within the field of system administration.

\section{Limitations}
The population we refer to as administrators is inherently diverse in terms of responsibilities, education, and previous experience. Depending on the size of a company, administrators have different responsibilities and work either in isolation or in larger teams. Furthermore, the security requirements depend on the types of products and services a company offers. Also, there is no unified career path for administrators, and one must not necessarily have a degree or certificate of any kind to become an administrator. 
Because of all these aspects, the results are not generalizable and thus applicable other populations of administrators with different demographics or training. The participants in the online survey were mainly from Europe and the United States. In these regions, technical staff like administrators are predominantly male which is why the sample was heavily biased in terms of gender. Due to our recruitment strategy for the quantitative study, the sample potentially suffered from self-selection bias, as was likely also due to the completion rate ($51.1\%$) of the survey. 
Regarding our questionnaire, we did not ask the participants about their current employment status. This could result in answers from people that worked as an administrator previously and are now in a different position. However, due to the mentioned self-selection bias we think that the participants are still somehow active in this area.
Also, we did not collect information about the systems and software, the administrators were in charge of. Because of this, we cannot report possible existing differences between, e.g., different operating systems or widespread versus niche software.
The analysis is based on self-reported data, and thus, participant reports are highly subjective. We have no reason to believe that social desirability and recall bias are uncommonly strong in the sample because the interviews and related work showed that administrators tend to admit that they do not know about everything \cite{Hrebec2001}. However, this must be taken it into account, especially when talking about risk, obstacle perception, and individual perception (e.g., P1). Finally, the qualitative interviews provided useful insights but did not reach saturation (cf.~\cite{Francis2010}). However, the potential lack of saturation is alleviated as the qualitative analysis was primarily used as an exploratory first step to build hypotheses. The answers to the free-text questions on the questionnaire did not bring up many new topics which make us confident that the most common real-world problems were covered. But, although several different issues were covered, we make no claim for completeness.

\section{Ethical Considerations}
Our institution located in central Europe does not have a formal IRB process for this type of study but has a series of guidelines to follow. According to these guidelines, we limited the collection of personal information as much as possible and collected data separately from contact information.
Furthermore, all the processes complied with the European General Data Protection Regulation (GDPR).
As the administration of services in a corporate environment is a sensitive topic, we did not collect detailed information about the companies' infrastructures. In addition, participants were explicitly given the chance to drop out at any time during the study. Finally, we emphasized the option to skip questions that participants preferred not to answer.

\section{Conclusion and Future Work}
This work contributes a mixed-methods study that revealed how administrators incorporate security updates in their daily work routines, what obstacles they experience, and what coping strategies they apply. We found that even experienced administrators find it hard to predict the direct consequences of applying an update and are heavily concerned about potential downtimes. Another interesting observation was that administrators often rely on information that is not provided by the (software) vendor but by online media or by their peers, who often face similar struggles. Among other things, the findings imply that there are aspects that vendors can influence such as, e.g., provide sufficient documentation or more granular updates, which can help to motivate administrators to update and support them in the update process.

Based on the insights presented in this paper, we recommend the following topics for future work: (1) Investigate current established formal processes and evaluate their effectiveness in supporting timely updates. (2) Create computer-supported solutions that enable better communication between administrators and in this way, enhance the transfer of knowledge. (3) Design and evaluate feasible tools that support situational awareness, e.g., by helping administrators to find out about relevant updates and provide them with the information they need.

\section{Acknowledgements}
We thank Karoline Busse for their support in the interview study and valuable input on the discussion. Also, we thank Jennifer Seifert for her help with the sociologial theory. Finally, we want to thank Matthew Smith, Frank Li and the anonymous reviewers that helped with their constructive feedback to improve this work.

%-------------------------------------------------------------------------------
\bibliographystyle{plain}
\bibliography{usenix2020_SOUPS.bib}

\section*{Appendix}
\begin{appendix}

\section{Questionnaire}
\label{app:quest_guideline}    
\subsection*{Information \& Consent}
Hello, we're Usable Security researchers from BLINDED and our mission is to make your challenges with
system updates easier. As a first step, we need to understand your experiences and struggles with
software updates in a corporate environment. We conducted interviews with seven colleagues of
you and condensed interesting themes. This short questionnaire will take about 10 minutes to
answer .
We know that your time is precious, which is why every tenth participant gets a 3D-print of a model
of her/his choice (max. 3x3x3cm and a reasonable model). If you are interested in this form of
compensation just leave us your email address in the commentary field at the end. This email
address will be stored separately from your answers and will only be used to communicate about
your compensation.
Please read all questions and instructions carefully. All of your answers will be checked, and your
survey may be rejected in the case of inconsistent answers. Your data will be collected and
processed in anonymized form, so that no connection to your person can be made.
You can stop participating in this study at any time. If you have any questions please contact us via
email.
\\

*1. I have read and understood the information provided above and consent to take part in this study.
\begin{itemize}
    \item I consent
    \item I do not consent
\end{itemize}

\subsection*{Demographics \& General}

*2. How old are you?

\textit{Text-input field}

3. What ist your gender?

\textit{Text-input field}

4. In what country do you work?

\textit{Text-input field}

*5. For how many years have you worked as a professional system administrator?

\textit{Text-input field}

\subsection*{Job information}
All of the questions on this page refer to a specific company. If you currently work as an administrator, please answer these questions about your current company. Instead, if you do not currently work as an administrator, please answer these questions about the last company at which you worked as an administrator. 

6. Is this company an IT company (software/hardware development, hosting, ISP, ...)?
\begin{itemize}
    \item Yes
    \item No
    \item Other (please specify): \textit{Text-input field}
\end{itemize}

7. Which of the following statements best describes your role in this company?
\begin{itemize}
    \item My primary responsibility was system administration
    \item My primary responsibility was not system administration, but I spent at least 20\% of my time on system administration
    \item My primary responsibility was not system administration, but I spent between 1\% and 19\% of my time on system administration
    \item I did not perform system administration at that company
\end{itemize}

8. In a few words, what would you consider as your main task in the company you are working at?

\textit{Text-input field}

9. What is your main task as a system administrator? If it is the same as in the previous answer, please
answer: same.

\textit{Text-input field}

10. What kind of systems do you administer?
\begin{itemize}
    \item Clients (e.g. workstations)
    \item Servers
    \item Mobile Clients (eg. tablet, smartphone)
    \item Other (please specify):
    \textit{Text-input field}
\end{itemize}

* 11. How big is the company you work at as a system administrator?
\begin{itemize}
    \item less than 10 employees
    \item up to 50 employees
    \item up to 250 employees
    \item more than 250 employees
\end{itemize}

12. Do you work in a team?
\begin{itemize}
    \item Yes, as a team leader
    \item Yes, as a team member
    \item No
    \item Other (please specify):
    \textit{Text-input field}
\end{itemize}

*13. What kind of job related education did you receive? (e.g. training, certificate, university)

\textit{Text-input field}

14. Which of the following statements best describes the security-related training you have received
concerning system administration?
\begin{itemize}
    \item I received security-related training for system administration at that company
    \item I did not receive security-related training for system administration at that company, but I have received such training at a previous company or school
    \item I have never received security-related training for system administration
\end{itemize}

\subsection*{Update Process}
 Please be reminded that we do not collect or store identifying information. In the following we are interested in your honest opinion.

15. Among all software updates you install for operating systems or any other software running on
systems, approximately what percentage do you estimate are \textbf{security} updates?

 \textit{Slider [0-100]}

16. Within your job as a system administrator, how much effort does it take you to keep the software on
your systems up-to-date?

 \textit{7-point Likert scale from ``1 - Nearly none'' to ``7 - Nearly all my capacity''}

17. What pre-deployment steps do you take before installing an update on a live system?
\begin{itemize}
    \item We install it on a test system.
    \item We install it on a small number of production systems before deploying it to all systems or to everyone.
    \item We install it directly on all production systems.
    \item Other (please specify):
    \textit{Text-input field}
\end{itemize}

18. What is the share of security related updates in relation to all updates (in \%)?

 \textit{Slider [0-100]}

19. Which of the following statements best describe the update process in the company?
\begin{itemize}
    \item There is a written document, that formally describes the steps in the update process.
    \item There is no written document but an informal guideline that is followed in the update process.
    \item There is no defined update process.
\end{itemize}

20. What is the typical time-span between the release of an update to the installation in a normal update process?

    \textit{Text-input field}

*21. Please indicate how often the following situations occur:

 \textit{Table of the following questions, with a 6-point Likert scale from ``1 - Never'' to ``5 - Always'' and the option ``Not sure'', per question.}

\begin{itemize}
    \item I feel that I am not sufficiently trained as an administrator.
    \item I think of work-
related consequences
when doing tasks that
have, in case of a failure,
an impact on my
company (e.g. downtime
of a service that
everyone uses).
\item I feel personally
responsible for keeping
the software on my
systems up-to-date.
\end{itemize}

22. Please indicate how often the following situations occur:

 \textit{Table of the following questions, with a 6-point Likert scale from ``1 - Never'' to ``5 - Always'' and the option ``Not sure'', per question.}

\begin{itemize}
    \item Stability considerations
hinder the installation of
an update.
    \item Risk considerations
hinder the installation of
an update.
\item Performance
considerations hinder the
installation of an update.
\item Priority/time
considerations hinder the
installation of an update.
\item Software updates are
prevented because of
other software (e.g.
dependencies).
\end{itemize}

23. Please indicate how often the following situations occur:

 \textit{Table of the following questions, with a 6-point Likert scale from ``1 - Never'' to ``5 - Always'' and the option ``Not sure'', per question.}

\begin{itemize}
    \item System stability
considerations are
irrelevant to the
installation of an update.
    \item The risk of
breaking dependencies
hinder the installation of
an update.
\item A patch that is known to
introduce errors hinder
the installation of an
update.
\item Downtimes caused by
the update
process hinder the
installation of an update.
\item Lack of information
about the changes an
update introduced hinder
the installation of an
update
\item Lack of education and
knowledge hinder the
installation of an update.
\end{itemize}

24. Please indicate how much you would agree/disagree with the statements.

 \textit{Table of the following questions, with a 7-point Likert scale from ``1 - Strongly disagree'' to ``4 - Undecided'' to ``7 - Strongly agree'', per question.}

\begin{itemize}
    \item Deploying security
updates in a timely
manner is important.
    \item Post-installation
problems in a live system
are only a minor concern
because they don't
happen frequently.
\item Users often install
software without the
knowledge of the
administrator.
\end{itemize}

25. Who makes the decision whether to update or not?
\begin{itemize}
    \item My team.
    \item Myself.
    \item My colleague(s).
    \item My supervisor.
    \item None of the above, please specify:
    \textit{Text-input field}
\end{itemize}

26. Please indicate how often the following situations occur:

 \textit{Table of the following questions, with a 6-point Likert scale from ``1 - Never'' to ``5 - Always'' and the option ``Not sure'', per question.}

\begin{itemize}
    \item I feel sufficiently trained
as an administrator.
    \item I can oversee the impact
an update would have on
our systems.
\item I can oversee the impact
of a failed update on our
system.
\item I can oversee the
security impact of
updates on our systems.
\end{itemize}

\subsection*{Source and Tools}
 
*27. What sources do you use to get information about current system updates?
\begin{itemize}
    \item Online publications/news (e.g. cnet.com, Hacker News, heise,...)
    \item Update management software
    \item (Software) Publisher newsletters
    \item External services (e.g. a company that is contracted to inform you)
    \item Mailing lists
    \item My users
    \item Other (please specify):
    \textit{Text-input field}
\end{itemize}

*28. What ist your main source to get information about current system updates?
\begin{itemize}
    \item Online publications/news (e.g. cnet.com, Hacker News, heise, ...)
    \item Update management software
    \item (Software) Publisher newsletters
    \item External services (e.g. a company that is contracted to inform you)
    \item Mailing lists
    \item My users
    \item Other (please specify):
    \textit{Text-input field}
\end{itemize}

29. Please explain your previous answer:

\textit{Text-input field}

\subsection*{Thank you!}

30. What do you think are the biggest obstacles in the update process?

 \textit{Text-input field}

31. Thank you for your participation! If you have any further comments for us: Don't hesitate to use the textbox!

 \textit{Text-input field}

32 . If you are interested in the 3D model print just leave your email in this field. We will only use this mail for the communication and will not link it to your answers. 

 \textit{Text-input field}

\section{Interview Guidelines}
\label{app:interview_guideline}
    \subsection*{Questions to explore}
        \begin{enumerate}
        \setlength\itemsep{0em}
            \item What does the update process look like?
            \item What obstacles are there?
            \item Who is involved?
            \item What is his/her personal experience and assessment?
        \end{enumerate}
    \subsection*{Introduction}
        \begin{enumerate}
        \setlength\itemsep{0em}
            \item How long has he/she done the job? What is the training? What is he/she doing on a daily basis?
            \item What are the systems?
            \item Does he/she work in a team?
            \item What is the scope of his/her actions?
            \item What tools are used?
        \end{enumerate}
    \subsection*{General update process (or a specific update story)}
        \begin{enumerate}
        \setlength\itemsep{0em}
            \item How does he/she come in contact with updates? 
            \item What is the time frame and the process?
            \item What tools are used?
            \item Who is involved?
            \item Where does the information come from?
        \end{enumerate}
    \subsection*{(Optional) A second story}
        \begin{enumerate}
        \setlength\itemsep{0em}
            \item How does he/she come in contact with updates? 
            \item What is the time frame and the process?
            \item What are the tools?
            \item Who is involved?
            \item Where does the information come from?
        \end{enumerate}
    \subsection*{End}
        \begin{enumerate}
        \setlength\itemsep{0em}
            \item Do they have a fixed update policy?
            \item Are there any feelings connected to new updates or the installation?
            \item Is he/she aware of potential impacts of not installed update/failures of the installation? (Are there stories?)
            \item Are there wishes concerning the process/tools?
            \item Questionnaire
        \end{enumerate}
        
%\newpage
\section{Demographics}
\input{tables/quant_demographics.tex}

\input{tables/interviewparticipants.tex}

\end{appendix}
%%%%%%%%%%%%%%%%%%%%%%%%%%%%%%%%%%%%%%%%%%%%%%%%%%%%%%%%%%%%%%%%%%%%%%%%%%%%%%%%
\end{document}

%% file: data.tex
\newcommand{\qualitativeParticipantsAmount}{seven}
\newcommand{\qualitativeParticipantsAmountDigit}{7}

\newcommand{\newCodesAfterLastInterview}{XXX}

\newcommand{\alphaBefore}{0.61}
\newcommand{\alphaAfter}{0.98}

\newcommand{\qntParticipantCountTotal}{141}

%% Scriptoutput Begin
\newcommand{\qntParticipantCount}{67}
\newcommand{\qntParticipantFailedAttentionCheck}{5}
\newcommand{\qntValidAgeAnswers}{66}
\newcommand{\qntMedianAge}{34.5}
\newcommand{\qntMeanAge}{34.6}
\newcommand{\qntStdAge}{7.9}
\newcommand{\qntMinAge}{22.0}
\newcommand{\qntMaxAge}{55.0}
\newcommand{\qntValidExperienceAnswers}{66}
\newcommand{\qntMedianExperience}{10.0}
\newcommand{\qntMeanExperience}{11.1}
\newcommand{\qntStdExperience}{7.0}
\newcommand{\qntMinExperience}{0.1}
\newcommand{\qntMaxExperience}{30.0}
\newcommand{\qntITCompany}{34}
\newcommand{\qntNonITCompany}{29}
\newcommand{\qntOtherITCompany}{4}
\newcommand{\qntSystemClients}{28}
\newcommand{\qntSystemServers}{63}
\newcommand{\qntSystemMobile}{14}
\newcommand{\qntSystemOther}{13}
\newcommand{\qntRoleFullAdminNoAnswer}{0}
\newcommand{\qntRoleFullAdmin}{50}
\newcommand{\qntRoleTwentyAdmin}{11}
\newcommand{\qntRoleOneAdmin}{6}
\newcommand{\qntRoleNoAdmin}{0}
\newcommand{\qntRoleAnswers}{67}
\newcommand{\qntRoleAnswersAvg}{1.343}
\newcommand{\qntSexMale}{58}
\newcommand{\qntSexFemale}{1}
\newcommand{\qntSexOther}{3}
\newcommand{\qntSexNA}{5}
\newcommand{\qntSexPlzCheckManuallyOthers}{TODO}
\newcommand{\qntTeamLeaderNoAnswer}{2}
\newcommand{\qntTeamLeader}{16}
\newcommand{\qntTeamMember}{39}
\newcommand{\qntTeamNo}{10}
\newcommand{\qntTeamOther}{0}
\newcommand{\qntTeamAll}{67}
\newcommand{\qntTeamAllAvg}{1.908}
\newcommand{\qntParticipantPrint}{23}
\newcommand{\qntCompanysizeLessTenNoAnswer}{0}
\newcommand{\qntCompanysizeLessTen}{4}
\newcommand{\qntCompanysizeUpFifty}{15}
\newcommand{\qntCompanysizeUpTwoFifty}{15}
\newcommand{\qntCompanysizeMoreTwoFifty}{33}
\newcommand{\qntCompanyAnswers}{67}
\newcommand{\qntCompanyAnswersAvg}{3.149}
\newcommand{\qntFeelnottrainedNoAnswer}{0}
\newcommand{\qntFeelnottrainedOne}{15}
\newcommand{\qntFeelnottrainedTwo}{33}
\newcommand{\qntFeelnottrainedThree}{14}
\newcommand{\qntFeelnottrainedFour}{2}
\newcommand{\qntFeelnottrainedFive}{3}
\newcommand{\qntFeelnottrainedAnswers}{67}
\newcommand{\qntFeelnottrainedAvg}{2.179}
\newcommand{\qntFeelnottrainedMean}{3.46600182406e+64}
\newcommand{\qntHOneSmallMean}{2.38235294118}
\newcommand{\qntHOneSmallStd}{0.77907115956}
\newcommand{\qntHOneBigMean}{1.9696969697}
\newcommand{\qntHOneBigStd}{1.10354113213}
\newcommand{\qntSufficcentlytrainedNoAnswer}{0}
\newcommand{\qntSufficcentlytrainedOne}{1}
\newcommand{\qntSufficcentlytrainedTwo}{7}
\newcommand{\qntSufficcentlytrainedThree}{13}
\newcommand{\qntSufficcentlytrainedFour}{29}
\newcommand{\qntSufficcentlytrainedFive}{17}
\newcommand{\qntSufficcentlytrainedAnswers}{67}
\newcommand{\qntSufficcentlytrainedAvg}{3.806}
\newcommand{\qntFearconsequencesNoAnswer}{2}
\newcommand{\qntFearconsequencesOne}{3}
\newcommand{\qntFearconsequencesTwo}{3}
\newcommand{\qntFearconsequencesThree}{12}
\newcommand{\qntFearconsequencesFour}{15}
\newcommand{\qntFearconsequencesFive}{32}
\newcommand{\qntFearconsequencesAnswers}{67}
\newcommand{\qntFearconsequencesAvg}{4.077}
\newcommand{\qntSrcOnlinePub}{54}
\newcommand{\qntSrcNewsletter}{41}
\newcommand{\qntSrcMaillist}{36}
\newcommand{\qntSrcSoftware}{45}
\newcommand{\qntSrcExtservice}{8}
\newcommand{\qntSrcUsers}{4}
\newcommand{\qntSrcOther}{17}
\newcommand{\qntSrcMainOnlinePub}{28}
\newcommand{\qntSrcMainNewsletter}{18}
\newcommand{\qntSrcMainMaillist}{11}
\newcommand{\qntSrcMainSoftware}{31}
\newcommand{\qntSrcMainExtservice}{3}
\newcommand{\qntSrcMainUsers}{4}
\newcommand{\qntSrcMainother}{12}
\newcommand{\qntOverseeconsequencesofupdateNoAnswer}{2}
\newcommand{\qntOverseeconsequencesofupdateOne}{0}
\newcommand{\qntOverseeconsequencesofupdateTwo}{13}
\newcommand{\qntOverseeconsequencesofupdateThree}{9}
\newcommand{\qntOverseeconsequencesofupdateFour}{29}
\newcommand{\qntOverseeconsequencesofupdateFive}{14}
\newcommand{\qntOverseeconsequencesofupdateAnswers}{67}
\newcommand{\qntOverseeconsequencesofupdateAvg}{3.677}
\newcommand{\qntOverseeconsequencesoffailedupdateNoAnswer}{2}
\newcommand{\qntOverseeconsequencesoffailedupdateOne}{0}
\newcommand{\qntOverseeconsequencesoffailedupdateTwo}{7}
\newcommand{\qntOverseeconsequencesoffailedupdateThree}{12}
\newcommand{\qntOverseeconsequencesoffailedupdateFour}{23}
\newcommand{\qntOverseeconsequencesoffailedupdateFive}{23}
\newcommand{\qntOverseeconsequencesoffailedupdateAnswers}{67}
\newcommand{\qntOverseeconsequencesoffailedupdateAvg}{3.954}
\newcommand{\qntTimelyupdatesNoAnswer}{0}
\newcommand{\qntTimelyupdatesOne}{0}
\newcommand{\qntTimelyupdatesTwo}{1}
\newcommand{\qntTimelyupdatesThree}{0}
\newcommand{\qntTimelyupdatesFour}{0}
\newcommand{\qntTimelyupdatesFive}{7}
\newcommand{\qntTimelyupdatesSix}{18}
\newcommand{\qntTimelyupdatesSeven}{41}
\newcommand{\qntTimelyupdatesAnswers}{67}
\newcommand{\qntTimelyupdatesAvg}{6.448}
\newcommand{\qntPerfomanceconsiderationsNoAnswer}{0}
\newcommand{\qntPerfomanceconsiderationsOne}{24}
\newcommand{\qntPerfomanceconsiderationsTwo}{27}
\newcommand{\qntPerfomanceconsiderationsThree}{7}
\newcommand{\qntPerfomanceconsiderationsFour}{9}
\newcommand{\qntPerfomanceconsiderationsFive}{0}
\newcommand{\qntPerfomanceconsiderationsAnswers}{67}
\newcommand{\qntPerfomanceconsiderationsAvg}{2.015}
\newcommand{\qntDowntimeshinderNoAnswer}{0}
\newcommand{\qntDowntimeshinderOne}{8}
\newcommand{\qntDowntimeshinderTwo}{22}
\newcommand{\qntDowntimeshinderThree}{13}
\newcommand{\qntDowntimeshinderFour}{18}
\newcommand{\qntDowntimeshinderFive}{6}
\newcommand{\qntDowntimeshinderAnswers}{67}
\newcommand{\qntDowntimeshinderAvg}{2.881}
\newcommand{\qntAfterinstallationproblemsNoAnswer}{0}
\newcommand{\qntAfterinstallationproblemsOne}{8}
\newcommand{\qntAfterinstallationproblemsTwo}{9}
\newcommand{\qntAfterinstallationproblemsThree}{8}
\newcommand{\qntAfterinstallationproblemsFour}{5}
\newcommand{\qntAfterinstallationproblemsFive}{12}
\newcommand{\qntAfterinstallationproblemsSix}{16}
\newcommand{\qntAfterinstallationproblemsSeven}{9}
\newcommand{\qntAfterinstallationproblemsAnswers}{67}
\newcommand{\qntAfterinstallationproblemsAvg}{4.313}
\newcommand{\qntUnknownswbyusersNoAnswer}{0}
\newcommand{\qntUnknownswbyusersOne}{18}
\newcommand{\qntUnknownswbyusersTwo}{9}
\newcommand{\qntUnknownswbyusersThree}{7}
\newcommand{\qntUnknownswbyusersFour}{8}
\newcommand{\qntUnknownswbyusersFive}{12}
\newcommand{\qntUnknownswbyusersSix}{6}
\newcommand{\qntUnknownswbyusersSeven}{7}
\newcommand{\qntUnknownswbyusersAnswers}{67}
\newcommand{\qntUnknownswbyusersAvg}{3.493}
\newcommand{\qntOverseeSecImpactNoAnswer}{4}
\newcommand{\qntOverseeSecImpactOne}{0}
\newcommand{\qntOverseeSecImpactTwo}{8}
\newcommand{\qntOverseeSecImpactThree}{16}
\newcommand{\qntOverseeSecImpactFour}{21}
\newcommand{\qntOverseeSecImpactFive}{18}
\newcommand{\qntOverseeSecImpactSix}{0}
\newcommand{\qntOverseeSecImpactSeven}{0}
\newcommand{\qntOverseeSecImpactAnswers}{67}
\newcommand{\qntOverseeSecImpactAvg}{3.778}
\newcommand{\qntPercentageofSecUpdatesMean}{59.3731343284}
\newcommand{\qntPercentageofSecUpdatesStd}{23.3211023057}
\newcommand{\qntLackOfInformationHinderNoAnswer}{1}
\newcommand{\qntLackOfInformationHinderOne}{15}
\newcommand{\qntLackOfInformationHinderTwo}{19}
\newcommand{\qntLackOfInformationHinderThree}{18}
\newcommand{\qntLackOfInformationHinderFour}{9}
\newcommand{\qntLackOfInformationHinderFive}{4}
\newcommand{\qntLackOfInformationHinderAnswers}{66}
\newcommand{\qntLackOfInformationHinderAvg}{2.508}
\newcommand{\qntNewErrorsHinderNoAnswer}{1}
\newcommand{\qntNewErrorsHinderOne}{10}
\newcommand{\qntNewErrorsHinderTwo}{18}
\newcommand{\qntNewErrorsHinderThree}{8}
\newcommand{\qntNewErrorsHinderFour}{11}
\newcommand{\qntNewErrorsHinderFive}{19}
\newcommand{\qntNewErrorsHinderAnswers}{67}
\newcommand{\qntNewErrorsHinderAvg}{3.167}
\newcommand{\qntInstallStagesTest}{36}
\newcommand{\qntInstallStagesPilot}{36}
\newcommand{\qntInstallStagesDirect}{22}
\newcommand{\qntInstallStagesOther}{9}
\newcommand{\qntInstallStagesTestAndPilot}{24}
\newcommand{\qntUpdateProcessFormalNoAnswer}{0}
\newcommand{\qntUpdateProcessFormal}{28}
\newcommand{\qntUpdateProcessInformal}{26}
\newcommand{\qntUpdateProcessNo}{13}
\newcommand{\qntUpdateProcessAll}{67}
\newcommand{\qntUpdateProcessAllAvg}{1.776}

%% file: tables/qual_phases.tex
\begin{table*}[h]
\small
\begin{tabularx}{\textwidth}{c c p{9.96cm}} 
\toprule 
\small{\textbf{Phase}} & \small{\textbf{Step}} & \small{\textbf{Obstacles}}\\ 
 \hline
  \rowcolor{Gray} Information & Becoming aware &\\ 
  \rowcolor{Gray}& Further details & Unsatisfying communication with the publisher*\\ 
  Deciding & Discussion & Stability (1); Risk of exploits (2); Performance (1); Priority (2); Missing expertise (1) \\
  \rowcolor{Gray}Preparation & Planning & Planning itself (3); Time of release (3); Communication (1); Missing documentation about the system and processes*  \\
  \rowcolor{Gray}& Backup &  \\ 
  \rowcolor{Gray}& Waiting for release &   \\ 
  \rowcolor{Gray}& Obtaining the patch & Missing patches (1) \\ 
  \rowcolor{Gray}& Automating & \\ 
  \rowcolor{Gray}& Informing users & \\ 
  Testing & Test system & Testing itself (1); Broken dependencies (4); Resources*;  Frequency of updates*\\
  & Pilot system \\
  & Problem solving with manufacturer  \\ 
  \rowcolor{Gray}Installation & Installation itself & Failure (2); Missing configuration options (1); Social pressure; System resources (2); Complexity (3); Missing tools (3); Heterogeneous system (6); Company structure (3); Impact on systems/users (2); Downtime (1); Installation method (manual/automatic) (1,1)  \\
  \rowcolor{Gray}& User interaction & Waiting for users (1) \\ 
  \rowcolor{Gray}& Reboot & Reboot itself (3); Old/Slow hardware (1) \\ 
  Post-Installation & Documentation  \\
  & Testing/Monitoring  \\
  & Troubleshooting &   \\ 
  & Reversing & Missing backup, failover, or redundancy*\\
 \bottomrule
 \\
\end{tabularx}
\caption{Overview of phases, steps, and obstacles. The number in brackets denotes the number of participants who mentioned this aspect in the interviews.  *Additional obstacles were
found through the questionnaire.}
\label{table:qual:phasesandsteps}
\end{table*}

%% file: tables/quant_observations.tex
\begin{table}[htp!]
    \centering
    \small
    \begin{tabularx}{\columnwidth}{p{0.75cm} p{7cm} }
     \toprule
         ID & Observation  \\
         \hline
         \addlinespace[0.2cm]
         & \textbf{Update Process and Information}\\
         U1& Online sources are an important source for administrators to get informed about updates. \\ 
         
         U2&Small companies have no formal update process. \\
         
         \addlinespace[0.2cm]
         & \textbf{Update Obstacles}\\
         O1& Performance considerations often hinder the installation of an update. \\
        
         O2& Update-caused downtimes delay the installation of an update (e.g., reboots)\\
        
         O3&Problems after the installation of an update on the live system are only a minor concern. \\
        
         O4& Lack of information hinder the update process. \\
        
         O5& User action (e.g., installing a software without the knowledge of the admin) can circumvent the update process and render it useless. \\
        
         \addlinespace[0.2cm]
        & \textbf{Human Factors}\\
         P1& Administrators of big companies feel sufficiently trained. \\
        
%         P2a & Administrators think that they often cannot oversee the consequences for an update.\\

 %        P2b & Administrators think that they often cannot oversee the consequences for a failed update.\\

         P2& Administrators think that timely updates are important. \\
         \addlinespace[0.2cm]
    \bottomrule
    \\
    \end{tabularx}
    \caption{Key observations based on qualitative results.}
    \label{tab:quant_observations}
\end{table}

%% file: tables/quant_eval.tex
\begin{table*}
    \centering
    \small
    \begin{tabularx}{\textwidth}{c p{9.46cm} c c c c c c c c} 
\toprule
 \textbf{ID} & \textbf{Statement} & 1 & 2 & 3 & 4 & 5 & * & \textbf{Plot} & \textbf{Median} \\ [1ex] 
 \hline
   \rowcolor{Gray}
  O1 & Performance considerations hinder the installation of an update. & \qntPerfomanceconsiderationsOne & \qntPerfomanceconsiderationsTwo & \qntPerfomanceconsiderationsThree & \qntPerfomanceconsiderationsFour & \qntPerfomanceconsiderationsFive  & \qntPerfomanceconsiderationsNoAnswer & \includegraphics[trim={2cm 12.5cm 2cm 15cm},clip,width=20pt]{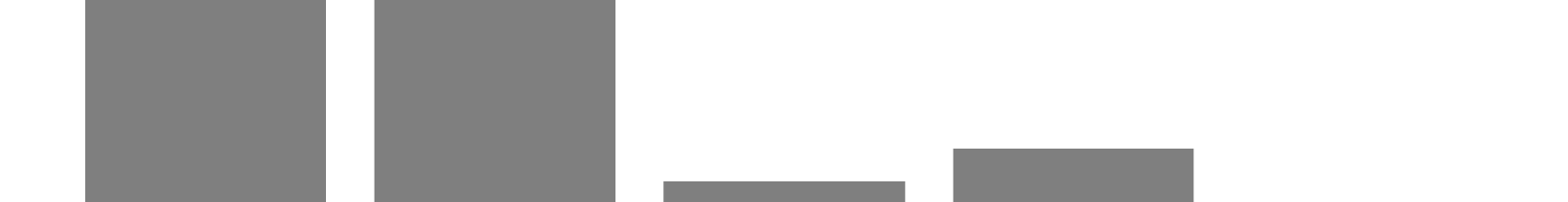} & 2 \\
 O2 & Downtimes caused by the update process hinder the installation of an update. & \qntDowntimeshinderOne & \qntDowntimeshinderTwo & \qntDowntimeshinderThree & \qntDowntimeshinderFour & \qntDowntimeshinderFive & \qntDowntimeshinderNoAnswer & \includegraphics[trim={2cm 12.5cm 2cm 15cm},clip,width=20pt]{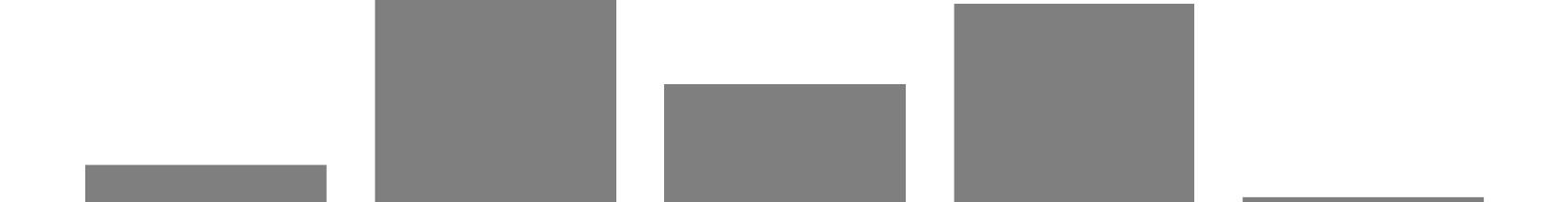} & 3 \\
   \rowcolor{Gray}
 O4 & A lack of information about the update hinder the installation of an update. & \qntLackOfInformationHinderOne & \qntLackOfInformationHinderTwo & \qntLackOfInformationHinderThree & \qntLackOfInformationHinderFour & \qntLackOfInformationHinderFive & \qntLackOfInformationHinderNoAnswer & \includegraphics[trim={2cm 12.5cm 2cm 15cm},clip,width=20pt]{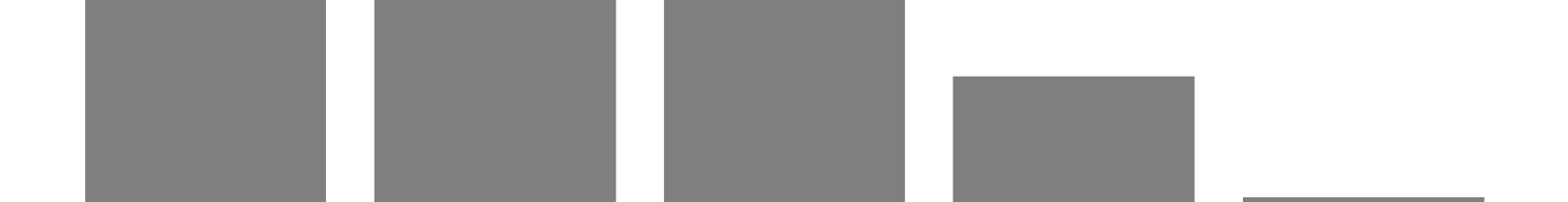} & 2\\
 P1 & I feel sufficiently trained as an administrator. & \qntSufficcentlytrainedOne & \qntSufficcentlytrainedTwo & \qntSufficcentlytrainedThree & \qntSufficcentlytrainedFour & \qntSufficcentlytrainedFive & \qntSufficcentlytrainedNoAnswer & \includegraphics[trim={2cm 12.5cm 2cm 15cm},clip,width=20pt]{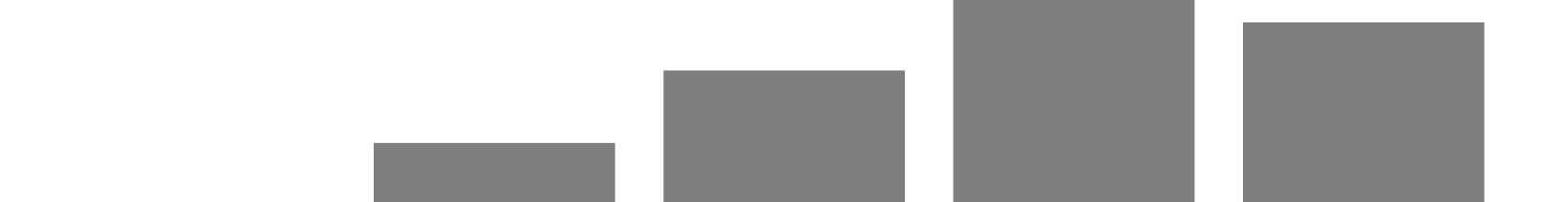} & 4 \\ 
 %P2 & I think of work-related consequences when doing tasks that have, in case of a failure, an impact on my company (e.g. downtime). & \qntFearconsequencesOne & \qntFearconsequencesTwo & \qntFearconsequencesThree & \qntFearconsequencesFour & \qntFearconsequencesFive & \qntFearconsequencesNoAnswer & \includegraphics[trim={2cm 12.5cm 2cm 15cm},clip,width=20pt]{images/PdfPictogramme/P2.pdf} & 4\\
   %\rowcolor{Gray}
 %P2a & I can oversee the impact an update would have on our systems. & \qntOverseeconsequencesofupdateOne & \qntOverseeconsequencesofupdateTwo & \qntOverseeconsequencesofupdateThree & \qntOverseeconsequencesofupdateFour & \qntOverseeconsequencesofupdateFive  & \qntOverseeconsequencesofupdateNoAnswer & \includegraphics[trim={2cm 12.5cm 2cm 15cm},clip,width=20pt]{images/PdfPictogramme/P3.pdf} & 4\\
 %P2b & I can oversee the impact a failed update would have on our systems. & \qntOverseeconsequencesoffailedupdateOne & \qntOverseeconsequencesoffailedupdateTwo & \qntOverseeconsequencesoffailedupdateThree & \qntOverseeconsequencesoffailedupdateFour & \qntOverseeconsequencesoffailedupdateFive  & \qntOverseeconsequencesoffailedupdateNoAnswer & \includegraphics[trim={2cm 12.5cm 2cm 15cm},clip,width=20pt]{images/PdfPictogramme/P4.pdf} & 4 \\
 \bottomrule
 \\
\end{tabularx}
\caption{Overview of the responses to statements regarding the frequency on a 5-point scale from ``1 - Never'' to ``5 - Always'' (* ``Not sure'') and their connection to the key observations.}
\label{table:quant:statementresponsesfivepoint}
\end{table*}

\begin{table*}
\small
\begin{tabularx}{\textwidth}{c p{8.46cm} c c c c c c c c c c} 
\toprule 
\textbf{ID} & \textbf{Statement} & 1 & 2 & 3 & 4 & 5 & 6 & 7 & * & \textbf{Plot} & \textbf{Median} \\[1ex] 
 \hline
 \rowcolor{Gray}
 O3 & Post-installation problems in a live system are only a minor concern because they don't happen frequently. & \qntAfterinstallationproblemsOne & \qntAfterinstallationproblemsTwo & \qntAfterinstallationproblemsThree & \qntAfterinstallationproblemsFour & \qntAfterinstallationproblemsFive & \qntAfterinstallationproblemsSix & \qntAfterinstallationproblemsSeven & \qntAfterinstallationproblemsNoAnswer & \includegraphics[trim={2cm 12.5cm 2cm 15cm},clip,width=20pt]{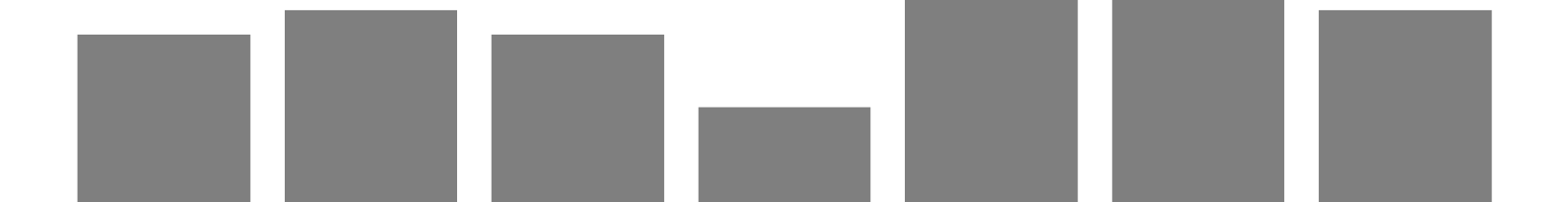} & 5 \\ 
 O5 & Users often install software without the knowledge of the administrator. & \qntUnknownswbyusersOne & \qntUnknownswbyusersTwo & \qntUnknownswbyusersThree & \qntUnknownswbyusersFour & \qntUnknownswbyusersFive & \qntUnknownswbyusersSix & \qntUnknownswbyusersSeven & \qntUnknownswbyusersNoAnswer & \includegraphics[trim={2cm 12.5cm 2cm 15cm},clip,width=20pt]{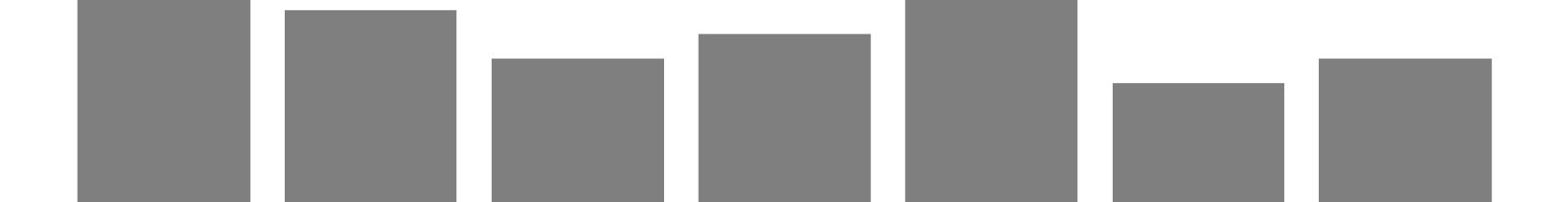} &  3\\
 \rowcolor{Gray}
 P2 & Deploying security updates in a timely manner is important. & \qntTimelyupdatesOne & \qntTimelyupdatesTwo & \qntTimelyupdatesThree & \qntTimelyupdatesFour & \qntTimelyupdatesFive & \qntTimelyupdatesSix & \qntTimelyupdatesSeven  & \qntTimelyupdatesNoAnswer & \includegraphics[trim={2cm 12.5cm 2cm 15cm},clip,width=20pt]{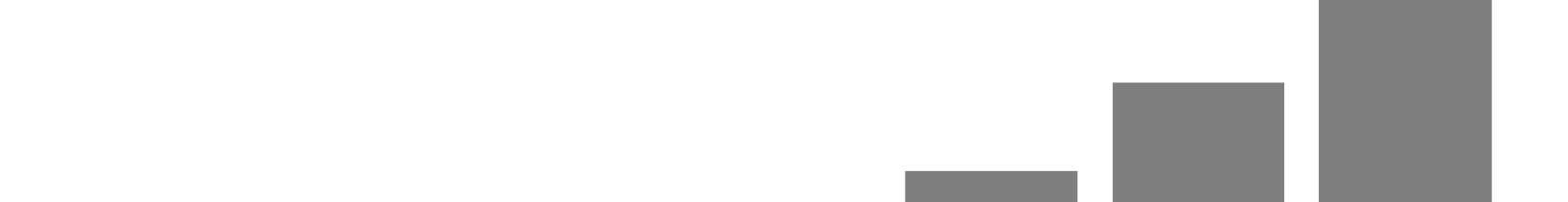} & 7 \\
[0.5ex] 
 \bottomrule
 \\
\end{tabularx}
\caption{Overview of the responses to statements regarding the attitude on a 7-point scale from ``1 - Strongly disagree'' to ``7 - Strongly agree'' (* ``Not sure'') and their connection to the key observations.}
\label{table:quant:statementresponsessevenpoint}

\end{table*}

%% file: tables/quant_timespan.tex
\begin{table}
    \centering
    \small
 \begin{tabularx}{\columnwidth}{ll}
    \toprule
        \small{\textbf{Interval}}& \\
        \hline
        \small{Hours to a day} &	\small{11} \\
        \small{Within a week} &	\small{19} \\
        \small{Within two weeks} &	\small{8} \\
        \small{Within one month} &	\small{11} \\
        \small{More than a month}&	\small{9} \\
        \small{No answer/no usable information (e.g., missing unit)} &	\small{11} \\
    \bottomrule
    \\
    \end{tabularx}
    \caption{Reported time intervals between the release of an update and deployment on all systems.}
    \label{tab:quant:timespan}
\end{table}

%% file: tables/quant_demographics.tex
\begin{table}[h!]
    \small
    \centering
 \begin{tabularx}{7cm}{p{2.2cm} p{4.5cm}}
    
         \multicolumn{2}{c}{\textbf{Survey demographic data}} \\
         \toprule
         \textbf{n} &  $\textbf{\qntParticipantCount}$\\
         \hline
         \textbf{Gender}& \qntSexFemale \quad Female\\
         & \qntSexMale \ \ Male\\
         & \qntSexOther \quad Other\\
         & 5 \quad Not specified\\
         \hline
         \textbf{Location}& 19 \ \  North America\\
         & 41 \ \ Europe\\
         & 7 \quad Rest of the world\\
         \hline
         \textbf{Age}&          \textbf{22 -- 55}\\
         Statistics & $md=34, mn=34.5, sd=7.8$\\
         \hline
         \textbf{Experience}&          \textbf{\qntMinExperience -- \qntMaxExperience } years\\
         Statistics & $md=\qntMedianExperience, mn=\qntMeanExperience, sd=\qntStdExperience$\\
        \hline
         
        \textbf{Company}
        & \qntITCompany \quad IT-related\\
        & \qntNonITCompany \quad Non IT-related\\
        & \qntOtherITCompany \ \ \quad Other\\
         \hline
         
         \textbf{Company Size}
        & \qntCompanysizeLessTen\ \ \quad $\leq10$\\
        & \qntCompanysizeUpFifty \quad $10 < x \leq 50$\\
        & \qntCompanysizeUpTwoFifty \quad $50 < x \leq 250$\\
        & \qntCompanysizeMoreTwoFifty \quad $>250$\\
         \hline
         
        % \textbf{Team}
        %& \qntTeamLeader \quad Team Leader\\
        %& \qntTeamMember \quad Team Member\\
        %& \qntTeamNo \quad No Team\\
        %\hline
         
          \textbf{Role}
        & \qntRoleFullAdmin \quad Full-time admin\\
        & \qntRoleTwentyAdmin \quad Not primary, but $>20\%$ of time\\
        & \qntRoleOneAdmin \ \ \quad Not primary, but $<20\%$ of time\\
         \hline
         
        \textbf{Administered} & \qntSystemClients \quad Clients\\
        \textbf{Systems }& \qntSystemServers \quad Servers\\
        & \qntSystemMobile \quad Mobile \\
        & \qntSystemOther \quad Other \\
     \bottomrule
     \\
    \end{tabularx}
    \caption{Demographic data from the online survey.   \label{tab:quant_demographic}}
\end{table}

%% file: tables/interviewparticipants.tex
\begin{table*}
    \small
    \centering
    \begin{tabularx}{\textwidth}{p{2cm} c c c c c p{5.05cm}}
    \multicolumn{2}{c}{\textbf{Interview demographic data}}  & & & & & \\
    \toprule
    \textbf{Pseudonym} & \textbf{Gender} & \textbf{Position/Task} & \textbf{Age} & \textbf{Exp. (Years)} & \textbf{Team size} & \textbf{Supervised Machines} \\
    \hline
    \rowcolor{Gray} Markus & M & Administrator & 25--35 & 6 & 7 & 300--350 clients, 150 virtual servers\\
    Lorenz & M & Update management & 25--35 & 2  & n/a & 5 servers\\
    \rowcolor{Gray} Cyril & M & Administrator & 25--35 & 6  & 15 & 10,000 virtual, ca. 100 physical\\
    Milan & M & Help desk & 25--35 & 2.5  & 12 & 600 clients, number of servers\\
    \rowcolor{Gray} Zelko & M & Administrator & 25--35 & 10  & 2 & 16 physical, 35 virtual, 80 clients\\
    Alexander & M & Update management & > 35 & 23  & 5 & 26 physical, 170 instances\\
    \rowcolor{Gray} Julian & M & Management & > 35 & 29  & 20 & n/a\\
    \bottomrule
    \end{tabularx}
    \caption{Interview participants.}
    \label{tab:interview_participants}
\end{table*}